\documentclass[a4paper]{jpconf}
\usepackage{graphicx}

\usepackage{amsmath,amssymb}
\usepackage{hyperref}
\newcommand{\unit}[1]{\ {\rm #1}}

\def\gagamma{g_{a\gamma}}

\begin{document}
\title{DANCE: Dark matter Axion search with riNg Cavity Experiment}

\author{Yuta~Michimura$^1$, Yuka~Oshima$^1$, Taihei~Watanabe$^1$, Takuya~Kawasaki$^1$, Hiroki~Takeda$^1$, Masaki~Ando$^{1,2}$,\\ Koji~Nagano$^3$, Ippei~Obata$^3$, Tomohiro~Fujita$^{4,5}$}
\address{$^1$ Department of Physics, University of Tokyo, Bunkyo, Tokyo 113-0033, Japan}
\address{$^2$ Research Center for the Early Universe, University of Tokyo, Bunkyo, Tokyo 113-0033, Japan}
\address{$^3$ Institute for Cosmic Ray Research, University of Tokyo, Kashiwa, Chiba, 277-8582, Japan}
\address{$^4$ Department of Physics, Kyoto University, Sakyo, Kyoto, 606-8502, Japan}
\address{$^5$ D{\'e}partment de Physique Theorique and Center for Astroparticle Physics, Universit{\'e} de Gen{\`e}ve, Quai E.Ansermet 24, CH-1211 Gen{\`e}ve 4, Switzerland}
\ead{michimura@granite.phys.s.u-tokyo.ac.jp}

\begin{abstract}
We have proposed a new approach to search for axion dark matter with an optical ring cavity [{\em Phys. Rev. Lett.\/} \href{https://doi.org/10.1103/PhysRevLett.121.161301}{{\bf 121}, 161301 (2018)}]. The coupling of photons to axions or axion-like particles makes a modulated difference in the phase velocity between left- and right-handed photons. Our method is to measure this phase velocity difference with a ring cavity, by measuring the resonant frequency difference between two circular polarizations. Our estimation shows that the sensitivity to axion-photon coupling constant $\gagamma$ for axion mass $m \lesssim 10^{-10}$~eV can be improved by several orders of magnitude compared with the current best limits. In this paper, we present the principles of the Dark matter Axion search with riNg Cavity Experiment (DANCE) and the status of the prototype experiment, DANCE Act-1.
\end{abstract}

\section{Introduction}
Axion is a pseudo-scalar field that is originally proposed to solve the strong CP problem in QCD physics. String theory and supergravity also generically predict a plenitude of axion-like particles (ALPs) which can have a broad range of mass. The QCD axion and ALPs are well-motivated candidates of dark matter, and often collectively called ``axion".

Axion behaves as a coherent field $a(t)=a_0 \cos(mt+\delta_\tau(t))$ with its mass $m$ and a phase factor $\delta_\tau(t)$, and may couple to photons through a Chern-Simons interaction term $\gagamma a F_{\mu \nu} F^{\mu \nu}/4$, where $\gagamma$ is the coupling constant. This interaction gives periodic conversion between axions and photons under a background magnetic field. A wide variety of experiments and astronomical observations to search for axions via this conversion have been carried out, but no signal has been found so far (for a review, see, for example, Ref.~\cite{Review}).

The axion-photon interaction term also gives a phase velocity difference between left- and right-handed circularly polarized light. The phase velocity difference $\delta c = |c_{\rm L} - c_{\rm R}| = \delta c_0 \sin(mt+\delta_\tau(t))$ for a wavelength of light $\lambda=2 \pi/k$ is estimated to be
\begin{equation} \label{phasevelocity}
 \delta c_0 = \frac{\gagamma a_0 m}{k} = 1.8 \times 10^{-24} \left( \frac{\lambda}{1064 \unit{nm}} \right) \left( \frac{\gagamma}{10^{-12} \unit{GeV^{-1}}} \right) .
\end{equation}
Here, we assumed axion energy density equals local dark matter density, $\rho_a = m^2 a_0^2 /2 \simeq 0.3 \unit{GeV/cm^3}$. The axion oscillation frequency is given by $2.4 \unit{Hz} (m/(10^{-14} \unit{GeV}))$.

Recently, several novel methods have been proposed to measure this modulated phase velocity difference using carefully designed optical cavities~\cite{AxionInterferometry,DANCE,ADBC,NaganoAxion}. The phase velocity difference gives resonant frequency difference between two circular polarizations. The polarization flip upon mirror reflection have to be taken into account to design the optical cavities. The proposal in Ref.~\cite{AxionInterferometry} uses linear cavities with quarter-wave plates inside to preserve the polarization of light. The proposal in Ref.~\cite{DANCE} instead uses a bow-tie ring cavity to avoid losses from wave plates, and uses a double-pass configuration to eliminate displacement noises.

The modulated phase velocity difference also gives the modulation of the polarization angle of linearly polarized light. The ADBC experiment~\cite{ADBC} exploits birefringence of mirrors in a bow-tie cavity to measure the polarization angle. By tuning the birefringence with the incident angle to the mirrors, resonant detection to improve the sensitivity at higher mass region is also possible. More recently, the authors of Ref.~\cite{NaganoAxion} have proposed the use of linear cavities in interferometric gravitational wave detectors to measure the modulation of the polarization angle. The sensitivity of this scheme drops off at low axion mass regions, but has sensitivity peaks when the round-trip light travel time for the cavity length is odd-multiples of the axion oscillation period.

These laser interferometric searches can be done without a strong magnetic field, and have been shown to be effective to search for axion dark matter in the low mass region. In this paper, we review our Dark matter Axion search with riNg Cavity Experiment (DANCE) proposal, and present the status of the prototype experiment, DANCE Act-1.

\section{Optical ring cavity search for axion dark matter}
\begin{figure}[t]
  \begin{center}
    \includegraphics[width=0.95\hsize]{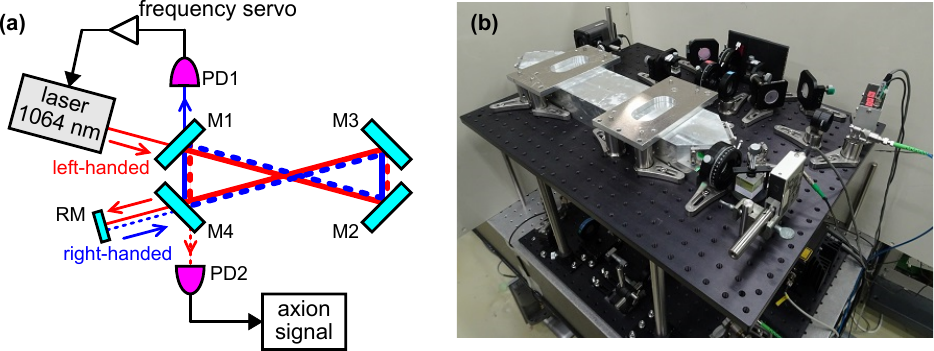}
    \caption{(a) Schematic of DANCE. The left-handed (right-handed) beam is drawn as red (blue) lines. The forward beam is drawn as solid lines and the beam reflected by RM is drawn as dotted lines. PD: photodetector. (b) Photo of the optical layout of DANCE Act-1.} \label{Setup}
  \end{center}
\end{figure}

Fig.~\ref{Setup}(a) depicts our approach to measure the resonant frequency difference between two circular polarizations using a double-pass bow-tie cavity. The left-handed circularly polarized beam from a single-frequency laser source is fed into the cavity, and the frequency of the laser is stabilized to the left-handed resonance. The transmitted light is then reflected back into the cavity in the opposite direction as right-handed polarization by a reflection mirror labeled RM. The signal proportional to the resonant frequency difference can be obtained from the photodetector labeled PD2, and in this signal we can search for axions.

The bow-tie configuration ensures the preservation of the polarization of light inside the cavity since the beam is reflected twice at both ends. The bow-tie configuration also cancels out the Sagnac effect since our cavity has zero loop-area. The double-pass configuration enables a null measurement of axion-photon coupling since the resonant frequency difference will be zero when the coupling is zero. Moreover, this differential measurement eliminates most of the environmental disturbances since the effects of cavity length fluctuations are common to both resonances~\cite{LV1,LV2}.

\begin{figure}[t]
  \begin{center}
    \includegraphics[width=0.75\hsize]{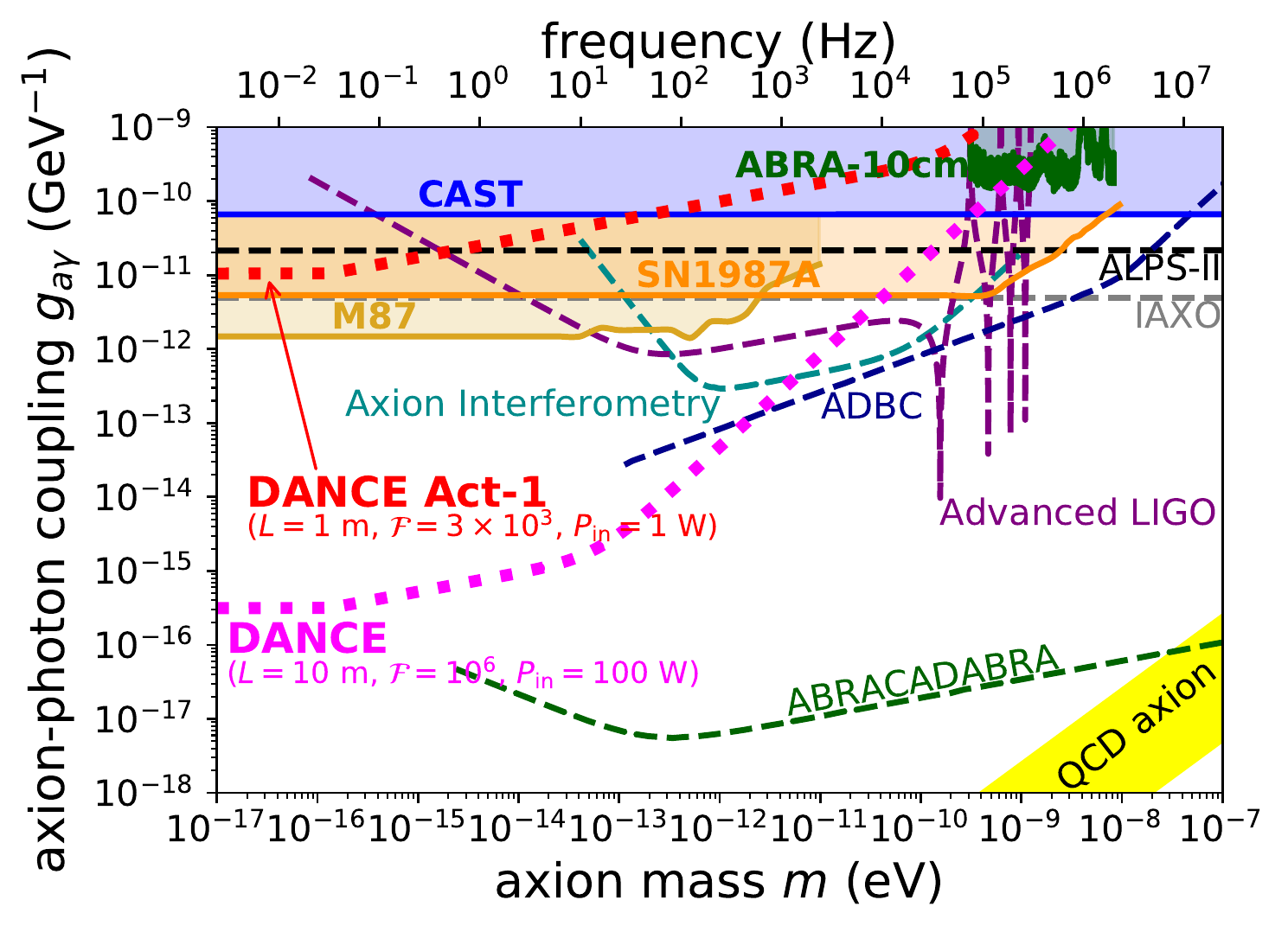}
    \caption{The sensitivity curves for the axion-photon coupling constant $\gagamma$. The expected shot noise limited sensitivity of DANCE and DANCE Act-1 with a year-long integration time are shown as dotted lines. The solid lines with shaded region are current bounds obtained from CAST~\cite{CAST} and ABRACADABRA-10cm~\cite{ABRA-10cm} experiments, and the astrophysical constraints from the gamma-ray observations of SN1987A~\cite{SN1987A} and the X-ray observations of M87 galaxy~\cite{M87}. The dashed lines are projected limits from ALPS-II~\cite{ALPS-II}, IAXO~\cite{IAXO} and ABRACADABRA broadband configuration with 5~T and 1~m$^3$~\cite{ABRACADABRA}. The expected limits from Axion Interferometry proposal with intra-cavity power of 1~MW~\cite{AxionInterferometry}, the expected integrated limits from ADBC experiment with intra-cavity power of 1~MW~\cite{ADBC}, and the expected sensitivity of Advanced LIGO with a scheme proposed in Ref.~\cite{NaganoAxion} are also shown with dashed lines for comparison.}  \label{Bounds}
  \end{center}
\end{figure}

Therefore, fundamental noise source of DANCE would be the quantum shot noise. The one-sided amplitude spectral density of the shot noise is given by
\begin{equation}
 \sqrt{S_{\rm shot}(\omega)} = \sqrt{\frac{\hbar \lambda}{4 \pi c P_{\rm in}} \left( \frac{1}{t_{\rm c}^2} + \omega^2 \right)},
\end{equation}
where $\omega$ is the fourier angular frequency and $P_{\rm in}$ is the incident power. The averaged storage time of the cavity $t_{\rm c}$ is given by $t_{\rm c}=L \mathcal{F}/(\pi c)$, where $L$ is the cavity round-trip length and $\mathcal{F}$ is the finesse. Assuming $L=10$~m, $\mathcal{F}=10^6$, and $P_{\rm in}=100$~W, we can reach $\gagamma \simeq 3 \times 10^{-16} \unit{GeV^{-1}}$ for $m \lesssim 10^{-16}$~GeV (see DANCE curve in Fig.~\ref{Bounds}). Here, we set $\lambda=1064$~nm and the integration time $T=1$~year.

The bending point of the sensitivity curve at $m=1.3 \times 10^{-16}$~eV is from the integration time being longer than the coherent timescale of axion oscillation at higher mass region. The coherent timescale is given by $\tau= 2\pi / (mv^2)$, where $v \simeq 10^{-3}$ is the local velocity of axion dark matter. The axion oscillation can be treated as coherent if $T \lesssim \tau$, and the signal-to-noise ratio (SNR) scales as $\sqrt{T}$ with this condition. If $T \gtrsim \tau$, the axion oscillation cannot be treated as coherent, and the SNR scales as $(T\tau)^{1/4}$. The bending point at $m=6.2 \times 10^{-14}$~eV is from the half linewidth of the cavity $f_{\rm c} = 1/(2 \pi t_{\rm c}) = c/(2 L \mathcal{F})$. The sensitivity at high mass region therefore cannot be improved by increasing $L$ or $\mathcal{F}$. For details of the sensitivity calculation, see Ref.~\cite{DANCE}.

\section{Prototype experiment}
Although cavity parameters discussed above are within the reach of state-of-the-art laser interferometry, we are planning to conduct a prototype experiment with moderate parameters to demonstrate the feasibility of our method and to investigate possible technical noises. Even with the cavity round-trip length of $L=1$~m, finesse of $\mathcal{F}=3 \times 10^3$, and input power of $P_{\rm in}=1$~W, we can reach the sensitivity beyond the CAST limit $\gagamma = 6.6 \times 10^{-11} \unit{GeV^{-1}}$ for $m \lesssim 10^{-12}$~eV (see DANCE Act-1 curve in Fig.~\ref{Bounds}).

The photo of the apparatus is shown in Fig.~\ref{Setup}(b). The bow-tie cavity is constructed from four mirrors rigidly fixed on a spacer made of aluminum. All the four mirrors are concave mirrors with 1~m radius of curvature. The length between mirrors M1 and M2 (or M3 and M4) is 45~cm, and that between M2 and M3 (or M4 and M1) is 4.7~cm. The $1/e^2$ beam radius at each mirror is calculated to be 0.39~mm. The input and output mirrors M1 and M4 are partially reflective mirrors with 99.9\% power reflectivity, which gives the designed finesse of $3 \times 10^3$. We use Pound-Drever-Hall method to obtain the error signal for the laser frequency servo.

At the time of writing, we have finished the assembly of the optics and optical characterization is underway. We expect to have the first data in 2020.

\section{Conclusion}
We have proposed DANCE to search for axion dark matter with a double-pass bow-tie cavity. Our scheme is highly insensitive to cavity length fluctuations, and no magnetic field is required. We showed that it is feasible to reach the sensitivity to the axion-photon coupling constant $\gagamma$ beyond the current bounds by several orders of magnitude for axion mass $m \lesssim 10^{-10}$~eV. A prototype experiment with a cavity round-trip length of 1~m is underway, and we look forward to having the first results in 2020.

\ack
We would like to thank Shigemi Otsuka and Togo Shimozawa for manufacturing the mechanical parts. This work was supported by JSPS KAKENHI Grant Numbers JP18H01224, JP17J01176 and JP17J09103.

\end{document}